\documentclass[%
reprint, superscriptaddress,
amsmath,amssymb,
aps, prb, 
]{revtex4-1}

\usepackage{graphicx}
\usepackage{dcolumn}
\usepackage{bm}
\usepackage{color,soul}

\begin{document}
\title{Nonreciprocal Second-Harmonic Generation in Few-Layer Chromium Triiodide}

\author{Wenshen Song}
\thanks{These authors contributed equally to this work.}
\affiliation{Department of Physics, Washington University in St.
Louis, St. Louis, MO 63130, USA}
\author{Ruixiang Fei}
\thanks{These authors contributed equally to this work.}
\affiliation{Department of Physics, Washington University in St.
Louis, St. Louis, MO 63130, USA}
\author{Li Yang}
\email{lyang@physics.wustl.edu} \affiliation{Department of
Physics, Washington University in St. Louis, St. Louis, MO 63130,
USA} \affiliation{Institute of Materials Science and Engineering,
Washington University in St. Louis, St. Louis, Missouri 63130,
USA}


\begin{abstract}
It is of fundamental importance but challenging to simultaneously
identify atomic and magnetic configurations of two-dimensional van
der Waals materials. In this work, we show that the nonreciprocal
second-harmonic generation (SHG) can be a powerful tool to answer
this challenge. Despite the preserved lattice inversion symmetry,
the interlayer antiferromagnetic order and spin-orbit coupling
generate enhanced SHG in PT-symmetric bilayer chromium triiodide
(CrI$_3$). Importantly, the in-plane polarization-resolved SHG is
sensitive to subtly different interlayer structures that cannot be
told by linear optical spectra. Beyond bilayer, we further predict
that the intensity and angle-resolved SHG can be employed to
identify both interlayer atomic and magnetic configurations of
trilayer CrI$_3$. Our first-principles results agree with
available measurements and show the potential of SHG as a
non-contacting approach to explore correlations between interlayer
structures and magnetic orders of emerging ultra-thin magnetic
materials.
\end{abstract}

\maketitle


\section{\label{intro}Introduction}

Antiferromagnetic (AFM) materials could represent the future of
spintronic applications due to many advanced features, such as
robustness against the magnetic-field perturbation, ultrafast
dynamics, and large magnetoresistance effects
\cite{baltz2018antiferromagnetic, nvemec2018antiferromagnetic,
wang2018very}. For example, experimental demonstrations of the
electrical switching and detection of the Neel order of AFM CuMnAs
open a new avenue towards memory devices based on antiferromagnets
\cite{Wadley2016}. However, simultaneously identifying both AFM
order and atomic structures is difficult for conventional
magnetometry approaches \cite{nvemec2018antiferromagnetic}. This
challenge is more serious for emerging two-dimensional (2D) van
der Waals (vdW) magnets. Although neutron diffraction plays a
major role in probing magnetic orders for bulk crystals
\cite{shull1951neutron}, it is not applicable for epitaxial or
exfoliated ultra-thin structures because of substrate effects. To
date magneto-optics effects, such as the magneto-optics Faraday
and Kerr effect, have been widely used for detecting the Neel
order of 2D magnets. \cite{Seyler2018, chen2019direct,
mak2019probing} Meanwhile, the terahertz radiation, which
predominantly interacts with low-energy excitations, is an
excellent tool for detecting the spin-wave excitation in magnetic
materials. Nevertheless, both probing tools cannot tell the subtle
interlayer structural information, which is, however, crucial for
studying vdW materials because their magnetic orders are strongly
correlated with interlayer configurations.
\cite{sivadas2018stacking, li2019pressure,prm19}

Second-harmonic generation (SHG) is a powerful tool to
discriminate the magnetic point or space groups that are
indistinguishable by the above diffraction methods
\cite{fiebig2000determination, fiebig2005second, Ju2009,
terrones2018}. Particularly, significant nonlinear optical (NLO)
responses were reported in bilayer AFM chromium triiodide
(CrI$_3$) \cite{Sun2019,Zhang2019}, indicating potentially unique
SHG of layered magnetic materials. Beyond studying CrI$_3$, which
exhibits a considerable magneto-optics effect
\cite{Sun2019,guo19mag,Zhang2019,wu2019physical}, SHG is also
capable in studying those AFM materials that have rather weak
magneto-optics effects \cite{wang2006ultrafast}. This advantage
broadens the applicable range of SHG. Correspondingly, a
systematical theoretical study of SHG and its relationship with
interlayer atomic and magnetic orders is crucial for understanding
available measurements and motivating further efforts to explore
complex symmetries of 2D vdW structures.

In this work, we focus on SHG of bilayer and trilayer CrI$_3$
using first-principles calculations. Despite the lattice inversion
symmetry, the interlayer AFM order and spin-orbit coupling (SOC)
of bilayer CrI$_3$ break both inversion and time-reversal
symmetries of magnetic space groups, leading to non-zero SHG
signals. We further find that the polarization-resolved azimuthal
SHG is determined by the space group of parent lattices and it can
be utilized to distinguish the subtle differences between
interlayer structures. For trilayer CrI$_3$, our calculation shows
that SHG can tell both interlayer atomic and magnetic orders,
giving rise to a powerful tool to efficiently explore symmetries
of ultra-thin 2D materials.

The remainder of this work is organized as follows: In Section
\ref{calc}, we present the calculation methods and simulation
details. In Section \ref{lattice}, the lattice structures of
CrI$_3$ and linear optical properties are introduced. In Section
\ref{origin}, we discuss the origin and mechanism of nonvanishing
SHG in PT-symmetric systems. In Section \ref{shg-bilayer}, the SHG
susceptibility of bilayer CrI$_3$ is obtained by first-principles
simulations. In Section \ref{pol-shg-bilayer}, we show the
polarization-resolved SHG and how to utilize it to distinguish
different interlayer structures of bilayer CrI$_3$. In Section
\ref{shg-trilayer}, we expand the SHG study to trilayer CrI$_3$
and show how to tell the interlayer atomic and magnetic
configurations. Finally, the results are summarized in the
conclusion section.

\section{\label{calc}Calculation details}

We calculate the structural and electronic properties of CrI$_3$
by density functional theory (DFT) within the generalized gradient
approximation (GGA) using the Perdew-Burke-Ernzerhof (PBE)
functional \cite{Perdew1996}, which is implemented in the Vienna
\emph{ab initio} simulation package (VASP) \cite{Kresse1996,
Kresse1999}. The vdW interactions are included through the DFT-D3
method with zero dampings \cite{Grimme2010, Grimme2011}. The
converged charge densities and wavefunctions are obtained in
plane-wave basis with an energy cutoff of 450 eV. SOC is included
in all simulations. To deal with localized $d$ orbitals of Cr
atoms, the DFT+U scheme is employed with $U=3$ eV
\cite{sivadas2018stacking}. Our main conclusions of SHG are not
sensitive to the value of U.

A k-point sampling of 32$\times$32$\times$1 is used to obtain
converged optical susceptibilities. The linear optical response is
calculated based on single-particle interband transitions
\cite{guo2004linear, wu2019physical, song, Wang2017}. DFT is known
for underestimating band gaps and neglecting many-electron
corrections, such as exciton effects \cite{rui20prb, chan19}.
However, in this study, we particularly focus on characters of
polarization dependencies of SHG according to symmetries of
materials. Therefore, we do not include excitonic effects, and the
band gap is corrected by rigidly shifting the DFT band gap to meet
the observed lowest-energy exciton peak \cite{wu2019physical}.
Within this approximation, the profiles of calculated NLO spectra
may not be accurate, but the polarization-symmetry dependence of
optical responses shall be reliable due to that electron-hole
(\emph{e-h}) interactions do not break any existing
structural/magnetic symmetry. For SHG calculations, we follow the
framework of perturbation theory based on the polarization
operator \cite{Hughes1996} by using the ArchNLO package.
\cite{song} We include 120 conduction bands for each layer of
CrI$_3$ for converged SHG spectra. Finally, we only consider
in-plane optical responses because of the local-field effect that
quenches off-plane electric
field\cite{ando1997excitons,machon2002ab,marinopoulos2003optical}.

\section{\label{lattice} Atomic structures and linear optical responses of bilayer AFM chromium triiodide}

\begin{figure}[hbt!]
\centering
\includegraphics[width=8.5 cm]{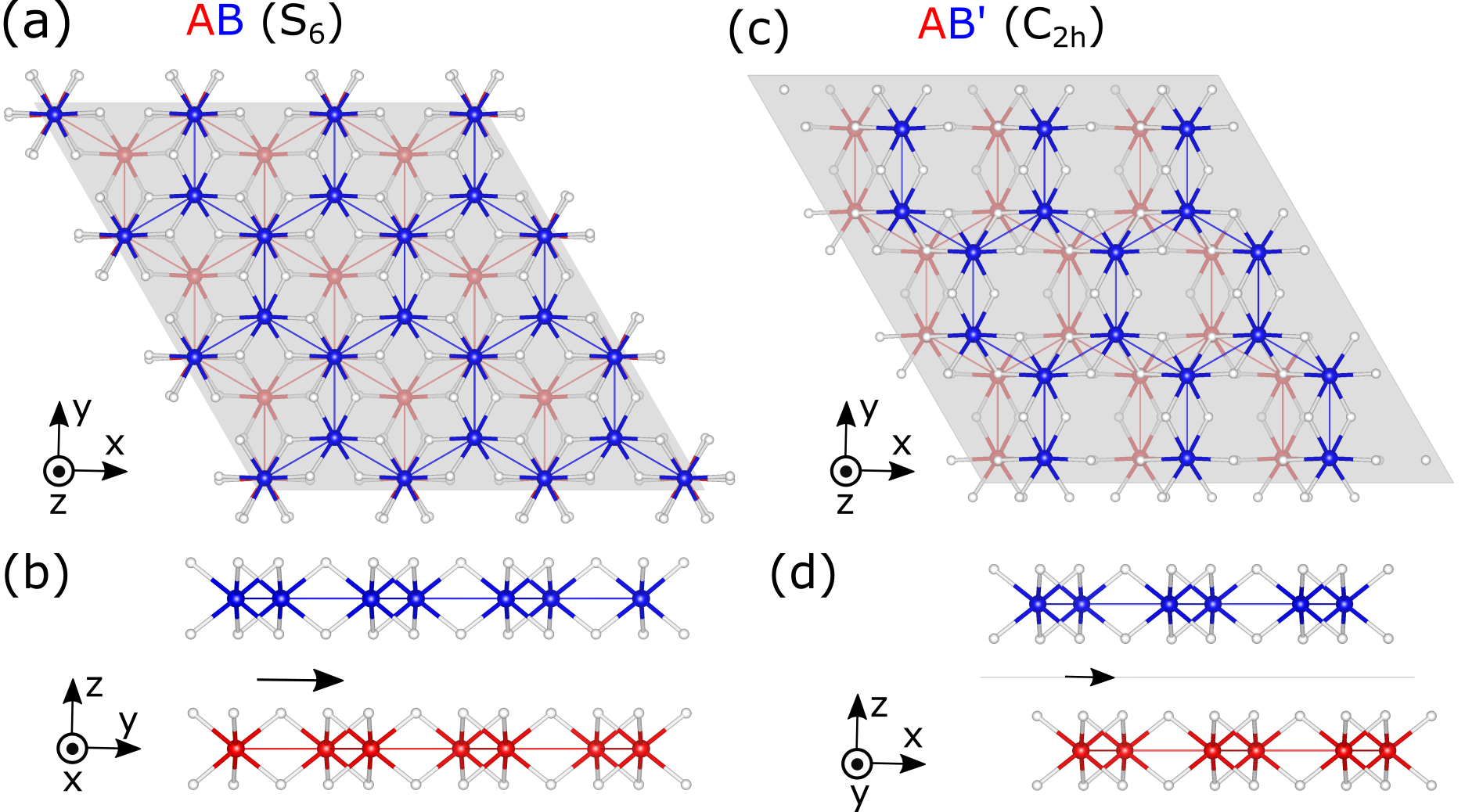}
\caption{\label{Fig 1} Top and side views of crystal structures of
bilayer CrI$_3$. (a) and (b) are the AB stacking structure with a
point group $S_6$, and it is corresponding to the rhombohedral
structure of bulk CrI$_3$ at low temperature. (c) and (d) are the
AB$^{\prime}$ stacking structure with a point group $C_{2h}$, and
it is corresponding to the monoclinic structure of bulk CrI$_3$ at
high temperature.}
\end{figure}

Bulk CrI$_3$ is a layered vdW magnetic material
\cite{handy1952structural}. Each layer of CrI$_3$ owns hexagonal
lattices in a $D_{3d}$ point group, in which magnetic Cr atoms
form a honeycomb structure and each Cr atom is surrounded by six I
atoms. A unit cell of bulk CrI$_3$ can be obtained by stacking
three monolayers following the ABC-Bernal configuration with the
same interlayer translation. There are two observed interlayer
structures \cite{mcguire2015cri3}. One is formed by a $[1/3, 1/3]$
interlayer shift, and it is observed at temperature below 210 K,
called the low-temperature (LT) structure. The other one is formed
by a $[1/3, 0]$ interlayer shift, and it is observed at
temperature above 210 K, called the high-temperature (HT)
structure \cite{mcguire2015cri3, weiji2019}.

The corresponding bilayer structures and symmetry groups based
these two bulk phases are presented in Figs. 1 (a)-(d), in which
the AB stacking style is from the bulk LT phase while the
AB$^{\prime}$ stacking style is from the bulk HT phase. In this
work, we mainly focus on the interlayer AFM order because of two
reasons: 1) the interlayer AFM order has been widely observed in
intrinsic 2D samples \cite{huang2017layer}; 2) the inversion
symmetry is preserved in FM bilayer CrI$_3$, and only the AFM
order exhibits non-zero SHG.

\begin{figure}[hbt!]
\centering
\includegraphics[width=8.5 cm]{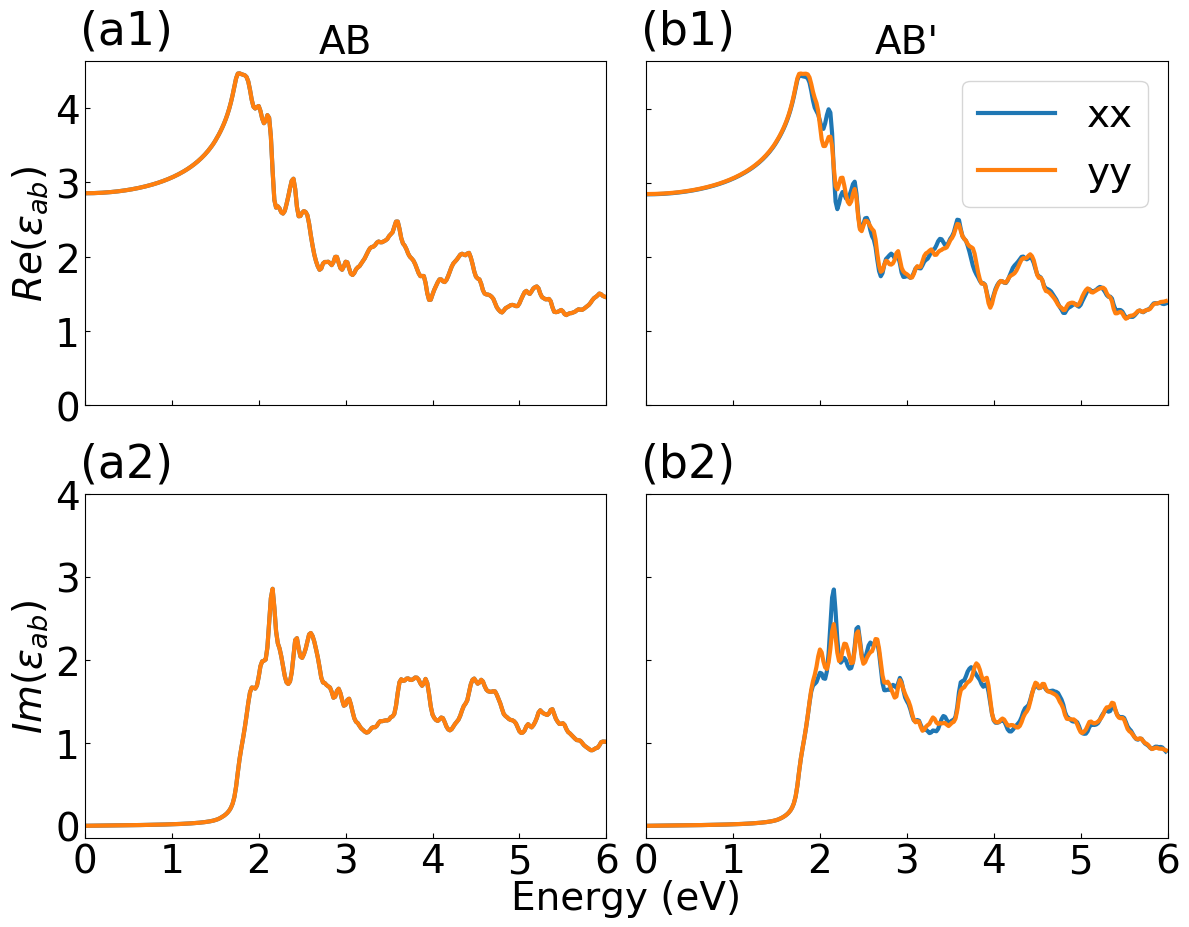}
\caption{\label{Fig 2} Real and imaginary parts of the in-plane
linear dielectric functions of AFM bilayer CrI$_3$. (a1) and (a2)
are those of the AB interlayer structure. (b1) and (b2) are those
of the AB$^{\prime}$ interlayer structure.}
\end{figure}

The DFT-calculated real and imaginary parts of the linear
dielectric function are plotted in Fig. \ref{Fig 2} for the AB and
AB$^{\prime}$ interlayer structures, respectively. Both AB and
AB$^{\prime}$ interlayer structures exhibit nearly identical
linear dielectric responses, except for a few minor differences.
The AB stacking shows a perfectly isotropic linear dielectric
function, where the $xx$ and $yy$ components are the same. This
isotropy is essentially decided by its $S_6$ point group. On the
other hand, the spectra of the AB$^{\prime}$ stacking exhibit a
minor anisotropy because of its lower-symmetric $C_{2h}$ point
group. However, this difference is too small to be detected in
practical linear optical spectra. Needless to say that the linear
dielectric function may not be an efficient way to tell interlayer
structures of bilayer CrI$_3$.

\section{\label{origin} Origin of SHG in parity-time symmetric AFM systems}

Before presenting the first-principles SHG simulation results, it
is necessary to introduce general expressions and discuss the
origin of non-zero SHG signals in interlayer AFM vdW systems,
which keep the parity-time (PT) symmetry.

Following previous works, \cite{Hughes1996,aversa1995nonlinear}
the SHG susceptibility
$\chi_{abc}^{(2)}\left(-2\omega;\omega,\omega\right)$ can be
generally expressed as
\begin{equation}
\label{eq:o1}
\begin{split}
\chi_{abc}^{(2)}\left(-2\omega;\omega,\omega\right)= \chi_{II}^{abc}\left(-2\omega;\omega,\omega\right)+\eta_{II}^{abc}\left(-2\omega;\omega,\omega\right)\\
+\sigma_{II}^{abc}\left(-2\omega;\omega,\omega\right).
\end{split}
\end{equation}

Specifically, the interband transitions at the same crystal
momentum $\boldsymbol{k}$ contribute to
\begin{equation}
\label{eq:o2}
\begin{split}
\chi_{II}^{abc}\left(-2\omega;\omega,\omega\right)=\frac{e^3}{\hbar^2}\sum_{nml}\int{\frac{d\boldsymbol{k}}{8\pi^3}\ \frac{r_{nm}^a{r_{ml}^br_{ln}^c}}{\omega_{ln}-\omega_{ml}}}\\
\times\left(\frac{f_{ml}}{\omega_{ml}-\omega}+\frac{f_{ln}}{\omega_{ln}-\omega\
}+\frac{2f_{nm}}{\omega_{mn}-2\omega}\right).
\end{split}
\end{equation}

The modulation of the linear susceptibility due to intraband
motions of electrons contributes to
\begin{equation}
\label{eq:o3}
\begin{split}
\eta_{II}^{abc}\left(-2\omega;\omega,\omega\right)=
\frac{e^3}{\hbar^2}\int\frac{d\boldsymbol{k}}{8\pi^3}\Bigg\{\sum_{nml}{\omega_{mn}r_{nm}^a\left\{r_{ml}^br_{ln}^c\right\}}\\
\left[\frac{f_{nl}}{\omega_{ln}^2\left(\omega_{ln}-\omega\right)}+\frac{f_{lm}}{\omega_{ml}^2\left(\omega_{ml}-\omega\right)}\right]\\
-8i\sum_{nm}\frac{f_{nm}r_{nm}^a}{\omega_{mn}^2\left(\omega_{mn}-2\omega\right)}\left\{\Delta_{mn}^br_{mn}^c\right\}\\
-\frac{2\sum_{nml}{f_{nm}r_{nm}^a\left\{r_{ml}^br_{ln}^c\right\}\left(\omega_{ln}-\omega_{ml}\right)}}{\omega_{mn}^2\left(\omega_{mn}-2\omega\right)}
\Bigg\}.
\end{split}
\end{equation}

Finally, the term describing the modification by the polarization
energy associated with interband motions contributes to
\begin{equation}
\label{eq:o4}
\begin{split}
\sigma_{II}^{abc}\left(-2\omega;\omega,\omega\right)=
\frac{e^3}{2\hbar^2}\int\frac{d\boldsymbol{k}}{8\pi^3}
\Bigg\{\sum_{nml}
\frac{f_{nm}}{\omega_{mn}^2\left(\omega_{mn}-\omega\right)}\\
\left[\omega_{nl}r_{lm}^a\left\{r_{mn}^br_{nl}^c\right\}-\omega_{lm}r_{nl}^a\left\{r_{lm}^br_{mn}^c\right\}\right]\\
+i\sum_{nm}\frac{f_{nm}r_{nm}^a}{\omega_{mn}^2\left(\omega_{mn}-2\omega\right)}\left\{\Delta_{mn}^br_{mn}^c\right\}\Bigg\}
\end{split}
\end{equation}

In Eq. \ref{eq:o1}$\sim$\ref{eq:o4}, we define the momentum matrix
element, $p_{ij}^a=\langle kj|\hat{p_a}|ki\rangle$, as the
transition between two states $i$ and $j$ at the same
$\boldsymbol{k}$ point, and the position matrix element is defined
by
$r_{nm}^a\left(k\right)=\frac{p_{nm}^a\left(k\right)}{im\omega_{nm}}$
if $n\neq m$ or 0 if $n= m$. Meanwhile,
$\left\{r_{ml}^br_{ln}^c\right\}\equiv\frac{1}{2}\left(r_{ml}^br_{ln}^c+r_{ml}^cr_{ln}^b\right)$
and
$\Delta_{mn}^b\equiv\frac{p_{mm}^b\left(k\right)-p_{nn}^b\left(k\right)}{m_e}$
($m_e$ is the electron mass).

It is known that the SHG response is zero in centro-symmetric
materials if time-reversal symmetry is preserved. This can be
understood from the above formulas, in which the
$r_{nm}r_{ml}r_{ln}$ and $r_{nm}r_{ml}\Delta_{ln}$ terms are odd
under inversion symmetry. As a result, their integrations over
reciprocal space are zero. However, if considering the magnetic
order, the inversion symmetry of the magnetic space group may be
broken even with a preserved lattice inversion symmetry, giving
hope to non-zero SHG.

\begin{figure}[hbt!]
\centering
\includegraphics[width=8.5 cm]{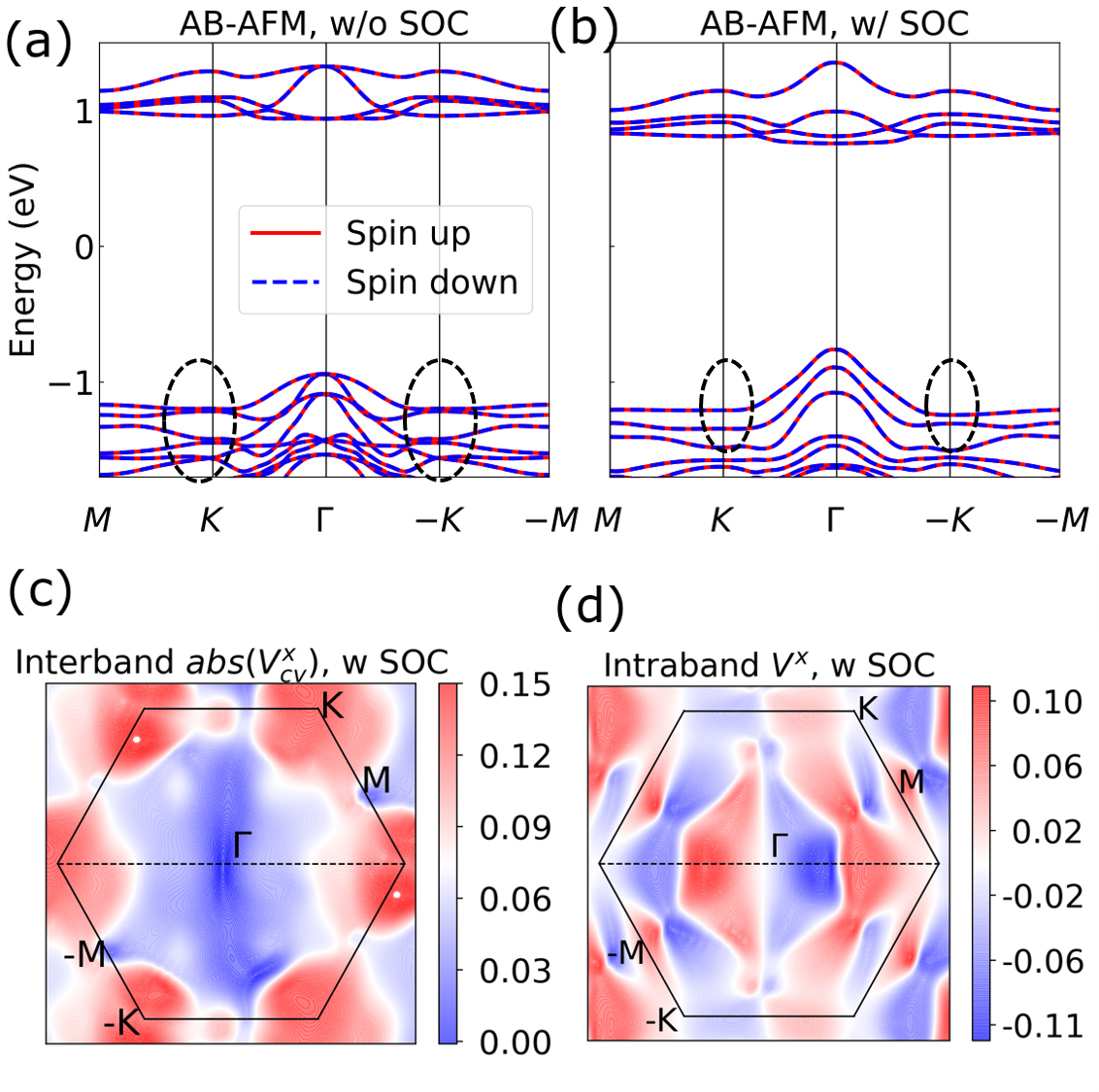}
\caption{\label{Fig 3} (a) Band structure of AFM AB bilayer
CrI$_3$ without SOC included. (b) is that with SOC included. The
Fermi level to set to be at the middle of the band gap. (c) The
distribution of the interband velocity matrix between the lowest
conduction and highest valence bands in reciprocal space. (d) The
distribution of the intraband velocity matrix of the lowest
conduction band in reciprocal space.}
\end{figure}

In both AB and AB$^{\prime}$ AFM bilayer CrI$_3$, the lattices own
the inversion symmetry. However, because the space inversion
operator cannot reverse the spin-degree freedom, the inversion
symmetry in the magnetic space group is broken. Interestingly,
this symmetry breaking itself does not guarantee non-zero SHG.
Particularly, due to PT symmetry, the spin-up and spin-down band
structures are degenerated, and they are symmetric in reciprocal
space because of the $SU(2)$ spin-rotation symmetry
\cite{mong2010afmtop}. For example, Fig.~\ref{Fig 3} (a) shows
this symmetric band structure of AFM AB bilayer CrI$_3$. As a
result, the SHG response is zero because the intraband group
velocity is odd in reciprocal space due to these symmetric band
structures.

Fortunately, SOC is known to be strong in CrI$_3$, and it breaks
the $SU(2)$ spin-rotation symmetry \cite{Zhang2019, Fei2019}. To
demonstrate the symmetry breaking, we plot the band structure with
SOC included for the AFM AB bilayer structure in Fig.~\ref{Fig 3}
(b). Black dotted circles are marked around $K$ and $-K$ points to
address the broken symmetric band structures by SOC. Moreover, as
shown in Figs. \ref{Fig 3} (c) and (d), the parity symmetries of
both interband and intraband velocity matrices, which are
calculated by first-principles simulations, are also broken with
SOC included. Because the strength of SHG susceptibility is
proportional to the integral of transition intensity and velocity
matrices in reciprocal space, these asymmetric velocity matrices
indicate non-zero SHG.

\section{\label{shg-bilayer} SHG susceptibility of bilayer AFM chromium triiodide}

\begin{figure}[hbt!]
\centering
\includegraphics[width=8.5 cm]{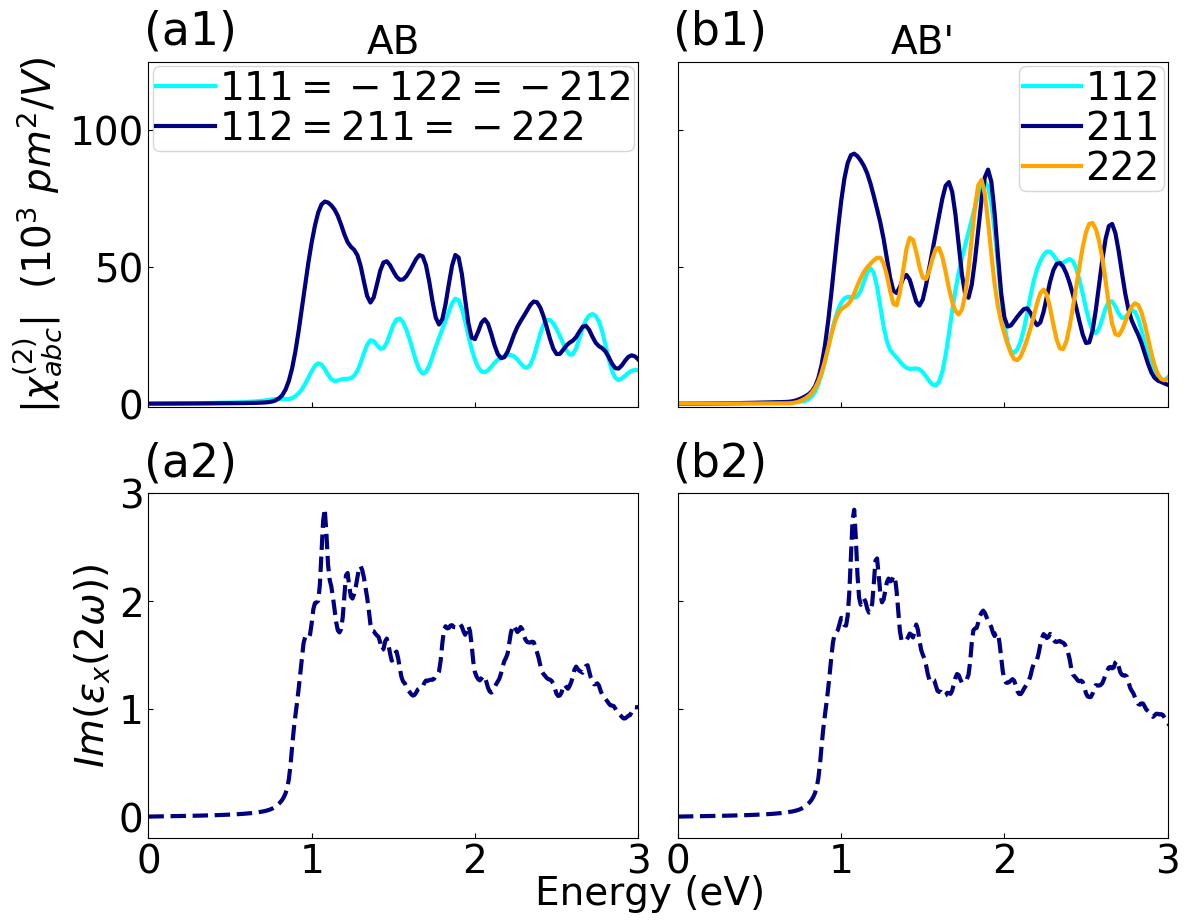}
\caption{\label{Fig 4} (a1) SHG spectra of the in-plane SHG
susceptibility (${|\chi}_{abc}^{(2)}|$) of AFM AB bilayer CrI$_3$.
(b1) are those of AFM AB$^{\prime}$ bilayer CrI$_3$. The
subscripts, 1, 2, and 3, denote the Cartesian coordinates x, y,
and z. (a2) and (b2) are the imaginary part of the double-resonant
dielectric function $\varepsilon_{x}\left(2\omega\right)$ of AB
and AB$^{\prime}$ bilayer CrI$_3$, respectively. }
\end{figure}

The in-plane components of SHG susceptibility tensor have been
calculated, and those nonzero components are plotted in
Figs.~\ref{Fig 4} (a1) and (b1) for AFM AB and AB$^{\prime}$
bilayer CrI$_3$, respectively. Unlike the similar linear optical
spectra in Fig.~\ref{Fig 2}, those NLO SHG spectra exhibit
significant differences between two interlayer structures.
Figure~\ref{Fig 4} (a1) presents that there are two non-zero
independent SHG spectra for the AB interlayer structure, and each
contains three degenerated components. The dark-blue line
represents the absolute SHG susceptibility elements of degenerated
$\chi_{112}^{(2)}$ = $\chi_{211}^{(2)}$ = ${-\chi}_{222}^{(2)}$.
while the cyan line represents those of degenerated
$\chi_{111}^{(2)}$ = ${-\chi}_{122}^{(2)}$ = $\chi_{212}^{(2)}$.
In the AB$^{\prime}$ stacking case shown in Fig.~\ref{Fig 4} (b1),
there are three non-zero independent SHG spectra,
$\chi_{112}^{(2)}$, $\chi_{211}^{(2)}$, and ${\chi}_{222}^{(2)}$
because of the lower symmetry.

To help analyze the spectra of SHG susceptibilities, we plot the
double-frequency linear optical absorption spectra
($Im(\varepsilon_x(2\omega)$)) in Figs.~\ref{Fig 4} (a2) and (b2)
of both interlayer structures. This is an approximation to only
consider two-photon processes with identical energy, reflecting
the double-photon resonance. \cite{song, Wang2015} Interestingly,
the profiles of double-resonant spectra and the significant
component of SHG spectra are similar. For example, the first
significant peak in the spectrum of $\chi_{112}^{(2)}$ is at 1.1
eV. It agrees well with the first peak of
$Im(\varepsilon_x(2\omega))$, as shown in Fig.~\ref{Fig 4} (a2).
Such a phenomenon indicates that main features (peaks) of SHG
spectra are dominated by double-resonance processes. This is
consistent with previous studies on transition-metal
dichalcogenides and hybrid halide perovskites\cite{song,
Wang2015}.

It is important to notice that the amplitude of SHG susceptibility
of bilayer PT-symmetric CrI$_3$ structures is significant. As
shown in Fig.~\ref{Fig 4}, their values can reach 7$\times$10$^4$
$pm^2/V$. These magnetic-ordering induced SHG signals are
comparable to those of monolayer MoS$_2$ (1$\times$10$^4$ $pm^2/V
\sim$ 6$\times$10$^4$ $pm^2/V$), which owns a non-centrosymmetric
structure \cite{Wang2017, 14prbGruning, li2013probing, zhao13prb,
malard13prb, trolle14prb}. Moreover, the SHG signals of bilayer
AFM CrI$_3$ are about one order of magnitude larger than those of
a hexagonal boron nitride (\emph{h}-BN) sheet (0.1$\times$10$^4$
$pm^2/V \sim$ 0.6$\times$10$^4$ $pm^2/V$) \cite{Wang2017,
14prbGruning, li2013probing}. This enhanced SHG agrees with recent
measurements of bilayer CrI$_3$ \cite{Sun2019}.

\section{\label{pol-shg-bilayer}Polarization-resolved SHG of bilayer chromium triiodide}

Although SHG spectra of AB and AB$^{\prime}$ interlayer structures
are different, it is not convenient to directly use them to
identify structures because this approach needs data of a wide
range of frequencies. A more efficient approach is to measure the
polarization-resolved SHG at a fixed frequency of the excitation
beam \cite{Sun2019, romijn2018automated, tBLG}. In the following,
we adopt the popular experimental setup and give the
angle-resolved SHG polarization of bilayer AFM CrI$_3$
\cite{Sun2019, Wang2017}. The response direction is set to be
parallelly (co-linearly) or perpendicularly (cross-linearly)
polarized with respect to the azimuthal polarization of incident
beam. Meanwhile, we keep these two directions rotating together
within the xy plane.

\begin{figure}[hbt!]
\centering
\includegraphics[width=8.5 cm]{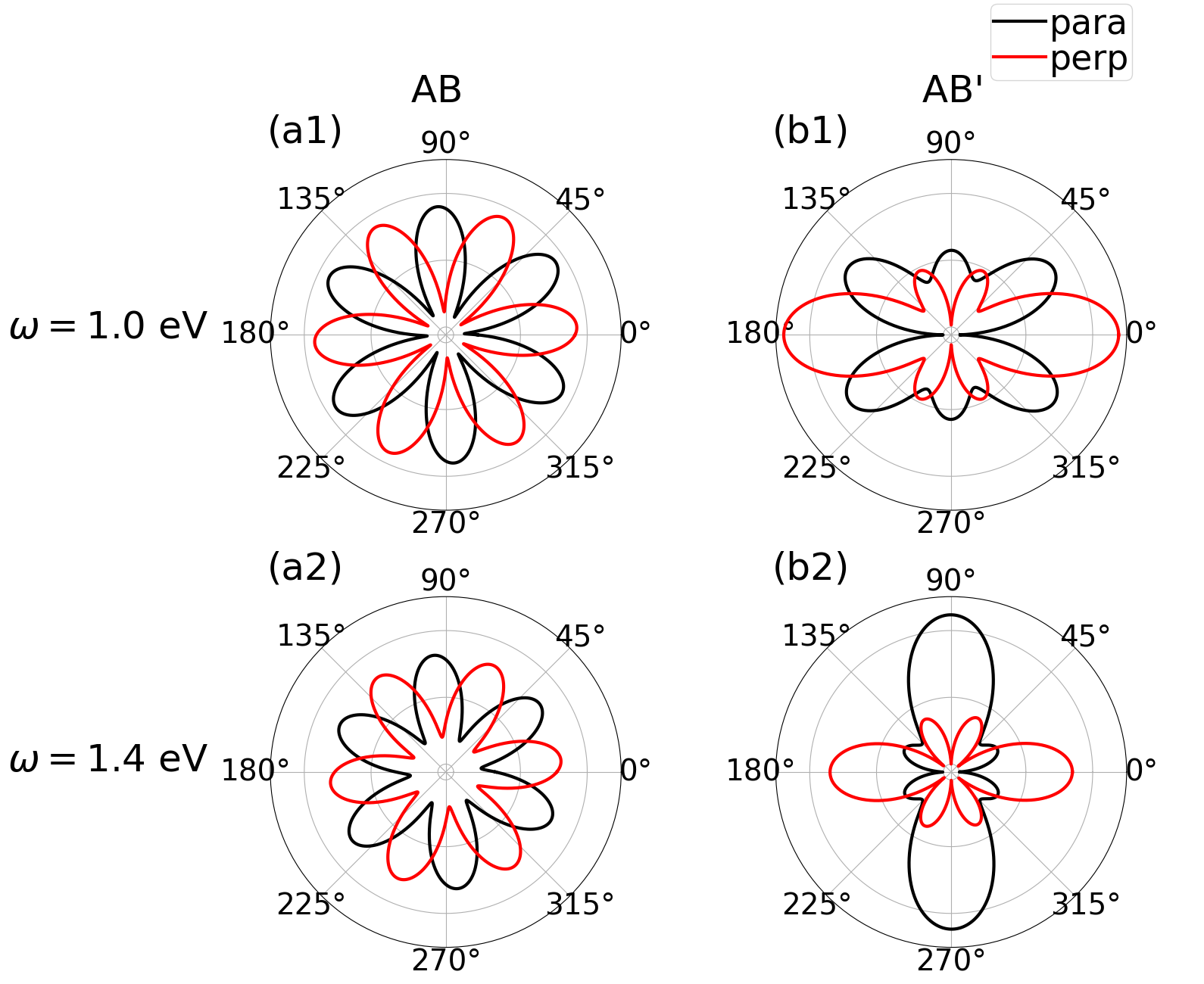}
\caption{\label{Fig 5} Polarization-resolved SHG of AFM bilayer
CrI$_3$. $\omega$ represents the energy of incident photons, and
para (perp) represents the parallelly (perpendicularly)
polarization component. (a1) and (a2) are those of the AB stacking
with incident photon at 1.0 eV and 1.4 eV, respectively. (b1) and
(b2) are those of the AB$^{\prime}$ stacking.}
\end{figure}

The electric field of incident light is given by
\begin{equation}
\label{eq:1}
\begin{cases}
    E_x=Ecos\theta\\
    E_y=Esin\theta,
\end{cases}
\end{equation}
in which $x, y$ donate laboratory coordinates, and $\theta$
represents the azimuthal rotational angle. $E_x$ and $E_y$ are
Cartesian components of the electric field of incident light.

In this work, we assume a normal incidence and focus on the
in-plane detection and excitation. The response of in-plane SHG
polarization is given by
\begin{equation}
\label{eq:2}
\begin{split}
\begin{cases}
    P_x=\chi_{111}^{(2)}E_x^2+2\chi_{112}^{(2)}E_xE_y+\chi_{122}^{(2)}E_y^2\\
    P_y=\chi_{211}^{(2)}E_x^2+2\chi
    _{212}^{(2)}E_xE_y+\chi_{222}^{(2)}E_y^2.
\end{cases}
\end{split}
\end{equation}
$\chi_{abc}^{(2)}$ presents components of SHG tensors, in which
the subscripts (1,2, and 3) donate $x$, $y$, and $z$. The first
subscript is the response direction, and the last two are the
excitation directions. $P_x$ and $P_y$ are induced polarizations
by excitation electric fields.

Finally, the parallel (perpendicular) SHG polarization can be
defined as
\begin{equation}
\label{eq:3}
\begin{cases}
    P_{\parallel}=P_xcos\theta+P_ysin\theta\\
    P_{\perp}=-P_xsin\theta+P_ycos\theta,
\end{cases}
\end{equation}
in which $\parallel$($\perp$) indicates the parallel
(perpendicular) polarization component.

Figure~\ref{Fig 5} shows the polarization dependence of SHG
responses for AB and AB$^{\prime}$ stackings of AFM bilayer
CrI$_3$ at two fixed frequencies ($\omega$=1.0 eV and 1.4 eV).
Unlike linear optical responses shown in Fig.~\ref{Fig 2}, the
in-plane polarization-resolved SHG is sensitive to the subtle
interlayer structures. In Figs.~\ref{Fig 5} (a1) and (a2) of the AB
stacking AFM bilayer CrI$_3$, both para and perp SHG signals
exhibit a 6-fold sunflower-like pattern. On the contrary, for the
AB$^{\prime}$ stacking, the SHG patterns exhibit a butterfly-like
two-fold mirror symmetry, as shown in Figs.~\ref{Fig 5} (b1) and
(b2).

These patterns are essentially decided by the symmetry groups of
interlayer structures. For the AB-type bilayer which owns a
high-symmetry $S_6$ point group, the 3-fold rotation symmetry
results in: $\chi_{111}^{(2)}=-\chi_{122}^{(2)}=\chi_{212}^{(2)}$
as well as $\chi_{112}^{(2)}=\chi_{211}^{(2)}=-\chi_{222}^{(2)}$
(indicated in Fig.~\ref{Fig 4} (a1)). We can substitute these
formulas into Eqs. \ref{eq:1}$\sim$\ref{eq:3}. The parallel and
perpendicular SHG susceptibilities are reduced to be
\begin{equation}
\label{eq:4}
\begin{cases}
    \chi_\parallel=\chi_{111}^{(2)}\cos{3\theta}+\chi_{112}^{(2)}\sin{3\theta}\\
    \chi_\bot={-\chi}_{111}^{(2)}\sin{3\theta}+\chi_{112}^{(2)}\cos{3\theta}.
\end{cases}
\end{equation}

Therefore, the absolute value of parallel and perpendicular
polarizations exhibit a 6-fold symmetry due to the $\cos{3\theta}$
and $\sin{3\theta}$ terms. Such a 6-fold SHG pattern has also been
observed in similarly three-fold-symmetry structures, such as
monolayer MoS$_2$ and $h$-BN \cite{Wang2017, tBLG, Kumar2013,
Malard2013, Li2013, Hsu2014}.

For the AB$^{\prime}$ stacking structure with a $C_{2h}$ symmetry,
the distinct non-zero in-plane elements of the SHG susceptibility
tensor are $\chi_{112}^{(2)}$, $\chi_{211}^{(2)}$, and
$\chi_{222}^{(2)}$, as shown in Fig.~\ref{Fig 4} (b1). In this
case, the parallel and perpendicular SHG polarization components
are reduced to be
\begin{equation}
\label{eq:5}
\begin{cases}
\chi_\parallel={(2\chi}_{112}^{(2)}+\chi_{211}^{(2)})\sin{\theta}\cos^2{\theta}+\chi_{222}^{(2)}\sin^3{\theta}\\
\chi_\bot={(-2\chi}_{112}^{(2)}+\chi_{222}^{(2)}){\sin}^2{\theta}\cos^2{\theta}+\chi_{211}^{(2)}\cos^3{\theta}.
\end{cases}
\end{equation}

These formulas of polarization components are complicated, leading
to the more anisotropic polarization-resolved SHG. However, the
absolute values are even-parity according to the angle ($\theta$),
resulting in a two-fold mirror symmetry, as shown in
Figs.~\ref{Fig 5} (b1) and (b2). These characteristic
mirror-symmetry patterns of the polarization-resolved SHG have
been also observed in monolayer group-IV monochalcogenides
\cite{Wang2017} because of their same point group.

These distinct features of polarization-resolved SHG in
Fig.~\ref{Fig 5} make it easy to distinguish interlayer structures
of bilayer AFM CrI$_3$. In fact, the similarly
polarization-resolved SHG shown in Figs.~\ref{Fig 5} (b1) and (b2)
were observed in fabricated bilayer CrI$_3$ with 1000-nm incident
light, confirming the AB$^{\prime}$ (HT) interlayer structure
\cite{Sun2019}. This observed AB$^{\prime}$ (HT) bilayer is
surprising because the structural phase transition temperature of
bulk CrI$_3$ is around 210 K \cite{mcguire2015cri3}, which is
substantially higher than the temperature ($\sim$5 K) measuring
SHG of bilayer structures. It will be valuable to explore the
fundamental reason for the preserved HT phase of ultra-thin
CrI$_3$ structures at low temperatures.

\begin{figure}
\centering
\includegraphics[width=8.5 cm]{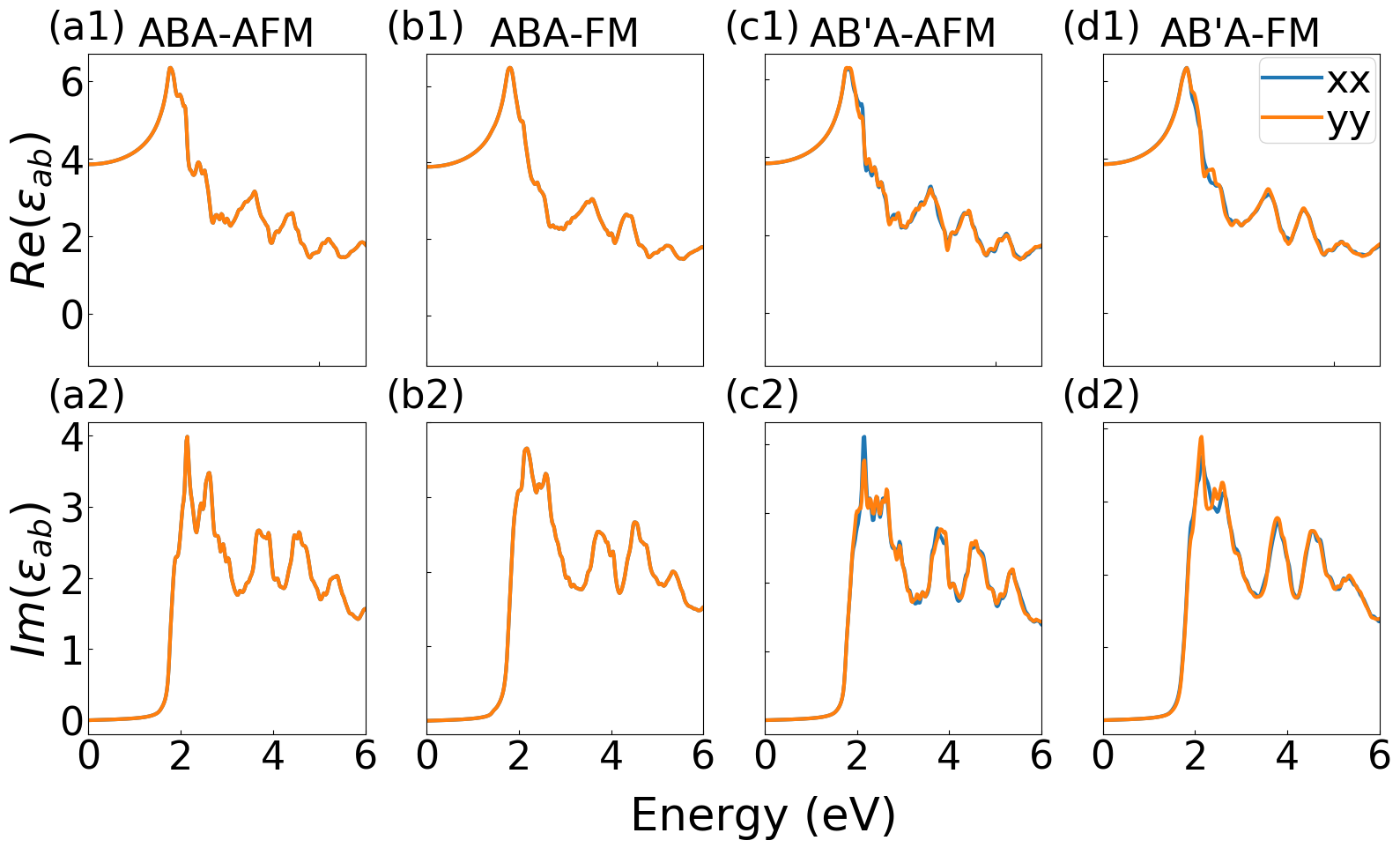}
\caption{\label{Fig 6} (a1) and (a2) are real and imaginary parts
of the in-plane linear dielectric functions of ABA-AFM trilayer
CrI$_3$, respectively. (b1) and (b2) are those of ABA-FM trilayer
CrI$_3$. (c1) and (c2) are those of AB$^{\prime}$A-AFM trilayer
CrI$_3$. (d1) and (d2) are those of AB$^{\prime}$A-FM trilayer
CrI$_3$.}
\end{figure}

\section{\label{shg-trilayer} Linear optical responses and SHG of trilayer chromium triiodide}

We focus on two stable interlayer configurations, i.e., the ABA
and AB$^{\prime}$A stacking styles of trilayer CrI$_3$. Unlike
bilayer, both FM and AFM orders break the inversion symmetry of
the magnetic group of trilayer CrI$_3$, resulting in non-zero SHG.
It is worth mentioning that, although most available measurements
observed an interlayer AFM order in few-layer CrI$_3$, a few
recent studies show that external factors, such as pressure, can
switch the interlayer magnetic ordering efficiently
\cite{li2019pressure}. Therefore, we will calculate SHG of both
interlayer AFM and FM orders of trilayer structures.

Figure~\ref{Fig 6} presents the real and imaginary parts of the
linear dielectric function of ABA and AB$^{\prime}$A
configurations with FM and AFM orders, respectively. Like those of
bilayer structures, the linear optical spectra of trilayer CrI$_3$
are nearly identical for different interlayer structures and
magnetic orders. Because of the lower symmetry of the
AB$^{\prime}$A interlayer structure, its linear spectra are
slightly anisotropic. Unfortunately, these minor differences may
not be significant enough to identify structural and magnetic
orders for trilayer CrI$_3$.

\begin{figure}[!]
\centering
\includegraphics[width=8.5 cm]{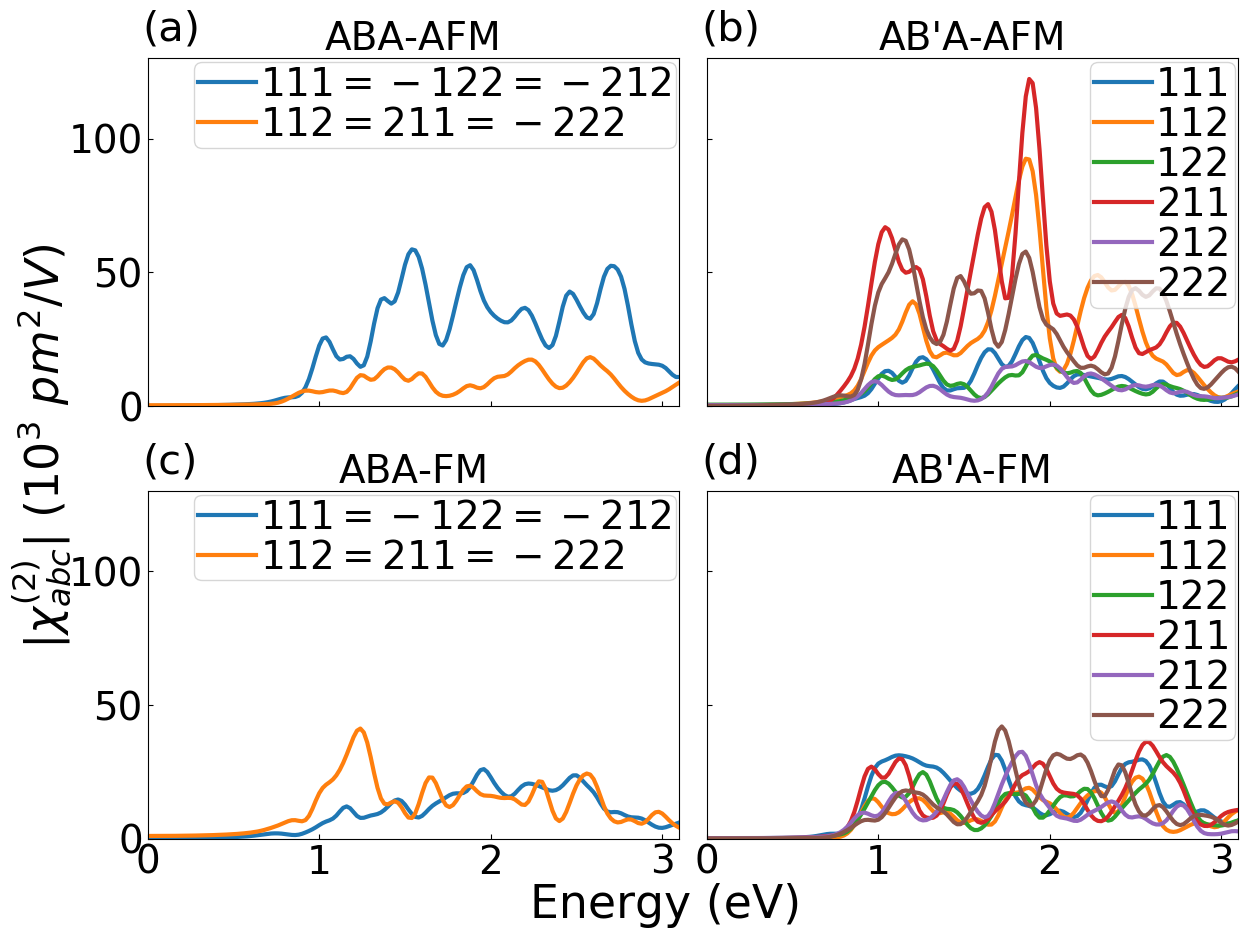}
\caption{\label{Fig 7} (a) SHG spectra of the in-plane SHG
susceptibility (${|\chi}_{abc}^{(2)}|$) of ABA-AFM trilayer
CrI$_3$. (b) Those of AB$^{\prime}$A-AFM trilayer CrI$_3$. (c)
Those of ABA-FM trilayer CrI$_3$. (d) Those of AB$^{\prime}$A-FM
trilayer CrI$_3$. The subscripts, 1, 2, and 3, denote the
Cartesian coordinates x, y, and z.}
\end{figure}

Figures ~\ref{Fig 7} (a)-(d) presents the SHG spectra of trilayer
CrI$_3$ with different interlayer structures and magnetic orders.
As expected, interlayer magnetic and atomic configurations
strongly affect SHG responses. The ABA stacking style has two
independent components for both FM and AFM, which are similar with
the bilayer case shown in Fig.~\ref{Fig 4} (a1), indicating a good
preservation of symmetries. The AB$^{\prime}$A stacking style has
six independent non-zero components, due to its lower symmetry.
Other than the different profiles of SHG spectra, we can observe
that the spectra of AFM structures have more significant peaks and
their average SHG intensities are also higher than those of FM
structures. This may provide an opportunity to identify magnetic
orders of trilayer CrI$_3$.

\begin{figure*}
\centering
\includegraphics[width=17 cm]{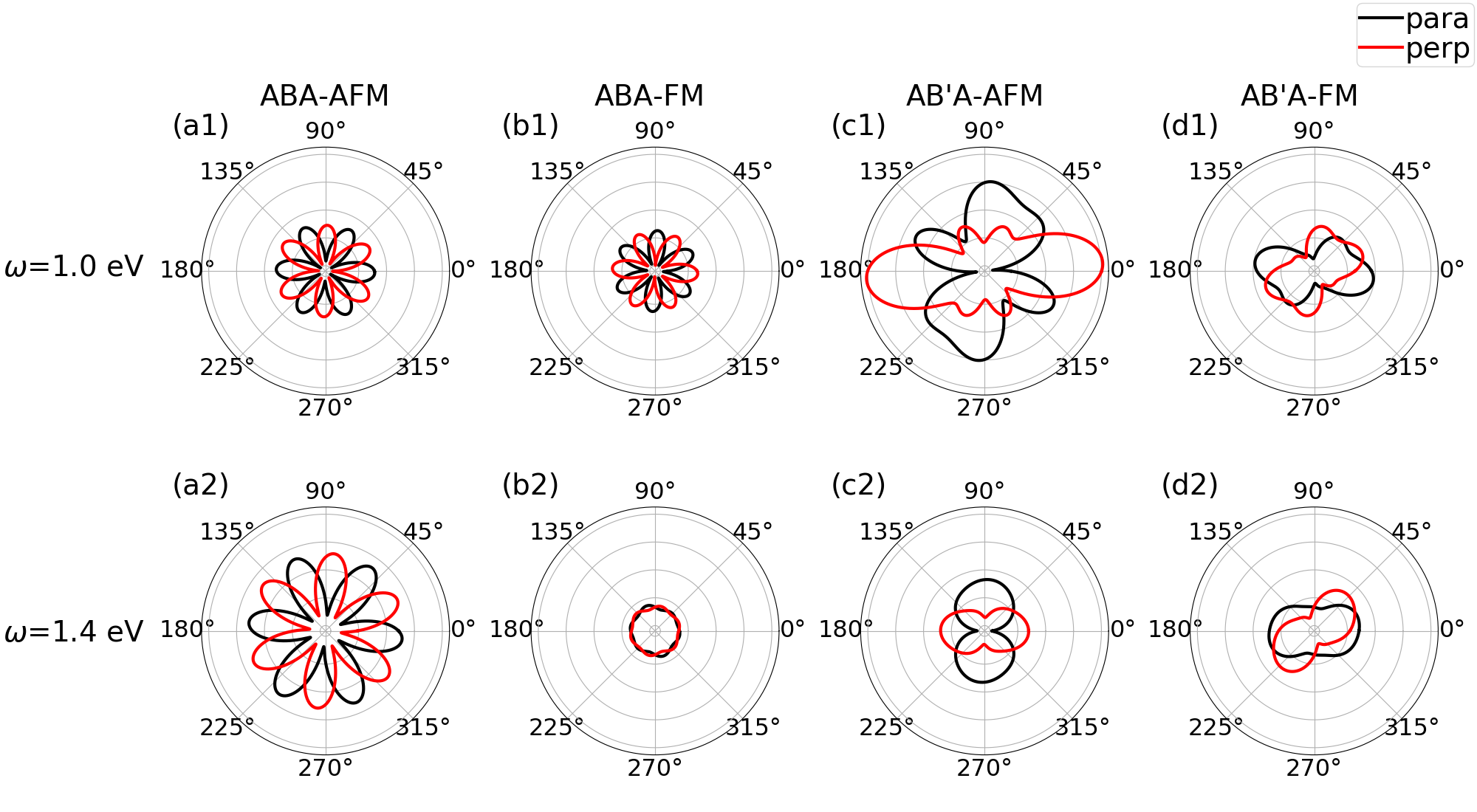}
\caption{\label{Fig 8} Polarization-resolved SHG of trilayer
CrI$_3$. $\omega$ represents energy of incident photons, and para
(perp) represents the parallelly (perpendicularly) polarization
component. (a1) and (a2) are those of the ABA-AFM trilayer CrI$_3$
with incident photon at 1.0 eV and 1.4 eV, respectively. (b1) and
(b2) are those of the ABA-FM trilayer CrI$_3$. (c1) and (c2) are
those of the AB$^{\prime}$A-AFM trilayer CrI$_3$. (d1) and (d2)
are those of the AB$^{\prime}$A-FM trilayer CrI$_3$.}
\end{figure*}

Following the same analysis stated in Section
\ref{pol-shg-bilayer}, we have further calculated the
polarization-resolved SHG patterns of trilayer CrI$_3$, which are
plotted in Fig.~\ref{Fig 8}. Two typical excitation
frequencies ($\omega$=1.0 eV and 1.4 eV) are considered in these
figures. In these angle-resolved cases, the SHG polarization is
more sensitive to the interlayer atomic structures than the
magnetic order. For example, for the ABA stacking style, both AFM
and FM orders exhibit a 6-fold sunflower-like pattern, which is
similar to the case of AB stacked bilayer. For the AB$^{\prime}$A
stacking style, both FM and AFM orders exhibit distorted
butterfly-like patterns, which are similar to the bilayer
AB$^{\prime}$ case but with a lower symmetry. We also notice that
the intensity of SHG polarizations of the AFM order is usually
stronger than those of the FM order. This is consistent with the
observations of Fig.~\ref{Fig 7}. As a result, the polarization
pattern of SHG is effective to tell the crystal structures while
its intensity may be useful to tell the magnetic order.

\section{\label{conclusion}Conclusion}
In summary, we have shown that the nontrivial AFM order and SOC
break the inversion symmetry and lead to enhanced SHG signals in
PT-symmetric bilayer CrI$_3$. Different patterns of
polarization-resolved azimuthal SHG can be utilized to distinguish
the AB and AB$^{\prime}$ interlayer structures. We further expand
this approach to discover both magnetic and interlayer structures
of trilayer CrI$_3$. The overall intensity of SHG signals can be
used to identify magnetic orders, and the polarization-resolved
SHG is effective to distinguish interlayer crystal structures. Our
calculation provides understandings of recent measurements and
sheds light on using nonlinear light-matter interactions to
explore atomic and magnetic structures of ultra-thin 2D vdW
materials.

\begin{acknowledgments}
This work is supported by the National Science Foundation (NSF)
CAREER grant No. DMR-1455346, NSF EFRI2DARE-1542815, and the Air
Force Office of Scientific Research (AFOSR) grant No.
FA9550-17-1-0304. The computational resources are provided by the
Stampede of Teragrid at the Texas Advanced Computing Center (TACC)
through XSEDE.
\end{acknowledgments}

%
\clearpage
\end{document}